\documentclass{sig-alternate-per}
\usepackage{bm}
\usepackage{framed}
\usepackage{dsfont}
\usepackage{graphicx}
\usepackage{caption}
\usepackage{xcolor}
 \usepackage{diagbox}
\usepackage{lettrine}
\usepackage{tipa}
 \usepackage{hyperref}
 \usepackage{framed}
 \usepackage{booktabs}
 \usepackage{caption}
\usepackage{url}
\usepackage{colortbl}
\usepackage{overpic}
\usepackage{amssymb}

\newtheorem{theorem}{Theorem}

\begin{document}
\conferenceinfo{MAMA}{2020, Boston, MA, USA	}
\title{Competitive Algorithms for Minimizing the Maximum Age-of-Information}
\numberofauthors{2}
\author{
\alignauthor
Rajarshi Bhattacharjee \\
       \affaddr{Indian Institute of Technology Madras}\\
       \affaddr{India}\\
       \email{brajarshi91@gmail.com}
\alignauthor
Abhishek Sinha \\
       \affaddr{Indian Institute of Technology Madras}\\
       \affaddr{India}\\
    \email{abhishek.sinha@ee.iitm.ac.in}
}
\maketitle
\begin{abstract} \label{abst}	
In this short paper, we consider the problem of designing a near-optimal competitive scheduling policy for $N$ mobile users, to maximize the freshness of available information uniformly across all users. Prompted by the unreliability and non-stationarity of the emerging 5G-mmWave channels for high-speed users, we forego of \emph{any} statistical assumptions of the wireless channels and user-mobility. Instead, we allow the channel states and the mobility patterns to be dictated by an omniscient adversary. It is not difficult to see that no competitive scheduling policy can exist for the corresponding throughput-maximization problem in this adversarial model. Surprisingly, we show that there exists a simple online distributed scheduling policy with a finite competitive ratio for maximizing the freshness of information in this adversarial model. Moreover, we also prove that the proposed policy is competitively optimal up to an $O(\ln N)$ factor. 
\end{abstract}


\section{introduction} \label{intro}
Apart from throughput, maximizing the freshness of information at the user-end is a principal design criterion for the emerging $5$G standards. The \emph{Age-of-Information} (AoI) is a newly proposed metric that captures the information-freshness in a quantitative fashion \cite{sun2019age}. However, the channel states and user-mobility are challenging to model and predict in $5$G-like non-stationary environments. 
This paper is concerned with the following question: Does there exist a scheduling policy that minimizes the maximum AoI across all users, irrespective of the channel dynamics and user-mobility patterns? Note that the question is considerably general, as it does not make any assumptions on either the channel-state statistics or the user-mobility, both of which may be dictated by an omniscient adversary in the worst case. In this paper, we affirmatively answer the above question by showing that a simple distributed greedy scheduling policy is competitively optimal up to an $O(\ln N)$ factor. 

Closely related to this work, in a recent paper \cite{banerjee2020fundamental}, we studied the problem of minimizing the \emph{average} AoI for $N$ \emph{static} users confined to a single cell within a similar adversarial framework. We showed that the greedy Max-Age (\textsf{MA}) policy is competitively optimal up to a factor of $O(N)$. In our previous paper \cite{srivastava2019minimizing}, we showed that the \textsf{MA} policy is optimal for minimizing the maximum AoI for static users in a single cell with a \emph{stochastic} channel state process. Within the stochastic framework, the paper \cite{kadota2018scheduling} proposes a Max-Weight scheduling policy, which is shown to be optimal up to a constant factor for the average AoI metric. Due to lack of space, we refer the reader to the book \cite{sun2019age} for a comprehensive introduction to this active area of research and an extensive bibliography. 
 
\textsc{\textbf{Contributions:}} Compared to the  previous works, this is the first paper to study the problem of minimizing the \emph{maximum} AoI for \emph{mobile} users in an \emph{adversarial framework}. Our main results are summarized and contrasted with that of \cite{banerjee2020fundamental} in Table \ref{summary}.
\begin{table*}
\caption{Summary of the results on the Competitive Ratios ($\eta$) in the adversarial framework} \label{summary}
\centering
\begin{tabular}{|l| ccc|>{\columncolor[gray]{0.8}}c|}
 \cmidrule(lr){1-1}\cmidrule(lr){2-4} \cmidrule(lr){5-5}
~~~~~~~~~~~\textbf{Metrics} & \textbf{Mobility} & \textbf{Upper Bound on $\eta$}  & \textbf{Lower Bound on $\eta$} & \textbf{Gap to optimality} \\
\midrule
Average Age \cite{banerjee2020fundamental}       &\textsf{No} & $O(N^2)$ & $O(N)$ & $O(N)$   \\
\midrule
Maximum Age (This paper) & \textsf{Yes} & $O(N)$ & $\Omega(\frac{N}{\ln(N)})$ & $O(\ln(N))$\\
\bottomrule
\end{tabular}
\end{table*} 
On the technical side, the proof of the achievability of Theorem \ref{upper_bound} differs from that of \cite{banerjee2020fundamental} in the way the ``\textsf{Max}-users" and ``super-intervals" are defined. This is essential because, due to user movements across multiple cells, the round-robin structure of scheduled users in \cite{banerjee2020fundamental} does not hold here anymore. Furthermore, the proof of converse in Theorem \ref{lower_bound} proceeds in a different way, making use of a  Maximal inequality similar to the Massart's lemma. 
\section{System Model} \label{adversarial}
A set of $N$ users move around in an area having a total of $M$ Base Stations (BS). The coverage areas corresponding to each BS (\emph{i.e.,} the cells) are disjoint. Time is slotted, and at any slot, a user can either stay in its current cell or move to any other $M-1$ cells of its choice. Our mobility model is considerably general, as it does not make any statistical assumptions on the speed or user movement patterns. At each slot, all BSs receive a fresh update packet for each user from an external source (\emph{e.g.,} a high-speed optical network). The fresh packets replace the stale packets in the BS buffers.  Each BS can beamform and schedule a downlink fresh packet transmission to only one user under its coverage area at a slot. The state of the channel for any user at any slot could be either \textsf{Good} or \textsf{Bad}. The BSs are assumed to be unaware of the current channel state conditions (\emph{i.e.,} no \textsc{CSIT}). If at any slot, a BS schedules a transmission to a user under its coverage having \textsf{Good} channel, the user decodes the packet successfully. Otherwise, the packet is lost. In the worst case, the states of the $N$ channels (corresponding to $N$ different users) and the user movements at every slot may be dictated by an \emph{omniscient adversary} (see, \emph{e.g.,} \cite{andrews2007routing}).

\textsc{\textbf{Cost Metric:}} In this paper, we are concerned with competitively optimizing the information freshness for all users. Accordingly, we define the $N$-dimensional state-vector $\bm{h}(t)$, where $h_i(t)$ denotes the length of the time interval prior to time $t$ before which the $i$\textsuperscript{th} user successfully received its most recent packet. The variable $h_i(t)$ is called the \emph{Age-of-Information} of the $i$\textsuperscript{th} user at time $t$ \cite{sun2019age}. Clearly, the graph of $h_i(t)$ has a saw-tooth shape that increases linearly with unit-slope until the $i$\textsuperscript{th} user receives a new packet, making $h_i(t)$ drop to $1$ at that slot. From that point onwards, $h_i(t)$ again continues increasing and repeats the saw-tooth pattern \cite{kadota2018scheduling}. 
The cost $C(t)$ at time $t$ is taken to be the maximum age among all users, \emph{i.e.,} $C(t)=\max_{i=1}^{N} h_i(t)$. The cumulative cost incurred over a time-horizon of length $T$ is defined as:
$\textsf{Cost}(T) = \sum_{t=1}^{T}C(t).$ 

\textsc{\textbf{Performance index:}}
As standard in the literature on online algorithms, we compare 
the performance of any online scheduling algorithm $\mathcal{A}$ against that of an optimal \emph{offline} scheduling algorithm \textsf{OPT} using the notion of competitive ratio $\eta^{\mathcal{A}}$,  defined as follows:
\begin{eqnarray}\label{comp_rat_def}
\eta^{\mathcal{A}} = \sup_{\bm \sigma}\bigg(\frac{\textrm{\textsf{Cost} of the online policy } \mathcal A \textrm{ on } \bm{\sigma}}{\textrm{\textsf{Cost} of offline OPT on } \bm{\sigma}}\bigg).	
\end{eqnarray}
In the above definition, the supremum is taken over all finite-length sequences $\bm \sigma$ denoting the  dynamic channel states and user locations per slot. Note that, while the online policy $\mathcal A$ has only  causal information, the policy \textsf{OPT} is assumed to be equipped with full knowledge (including the future) of the entire sequence $\bm \sigma.$ Our objective is to design an online scheduling policy $\mathcal{A}$ with the minimum competitive ratio.  
\section{Achievability} \label{upper_bound_section}
We consider the following distributed scheduling policy, called \textsf{Cellular Max-Age (CMA)}: At every slot, each BS $j$ schedules a transmission to the $i$\textsuperscript{th} user that has the maximum age $h_i(t)$ among all other current users in  BS $j$'s coverage area (ties are broken in an arbitrary but fixed order). 
Theorem \ref{upper_bound} below gives a performance bound for \textsf{CMA}, which is, quite surprisingly, independent of the number of BSs $M$. 
\begin{theorem} \label{upper_bound}
$\eta^{\textsf{CMA}} \leq 2N.$	
\end{theorem}

	
\textsc{Proof:} 
At any slot $t$, define the global  ``\emph{\textsf{Max}-user}" that has the highest age among all $N$ users (ties are broken in the same way as in the \textsf{CMA} policy). Note that the identity of the \textsf{Max}-user changes with time. However, by definition, the \textsf{CMA} policy continues to schedule the user corresponding to the current \textsf{Max}-user \emph{irrespective} of its locations until the transmission is successful. In the subsequent slot, a different user assumes the role of the \textsf{Max}-user, and the process continues. 

Let $T_i$ be the time slot at which a total of $i$ successful packet transmissions have been made exclusively by the \textsf{Max}-users. Let $\Delta_i \equiv T_i- T_{i-1}$ denote the length of the $i$\textsuperscript{th} \emph{super-interval}, defined as the time interval between the $i$\textsuperscript{th} and $i-1$\textsuperscript{th} successful transmissions by the \textsf{Max}-user. The super-intervals are contiguous and disjoint. Let the user $M_i$ be the \textsf{Max}-user corresponding to the $i$\textsuperscript{th} super-interval. 
As argued above, the user $M_i$ gets scheduled by the \textsf{CMA} policy persistently during the entire $i$\textsuperscript{th} super-interval of length $\Delta_i$, irrespective of its locations. Note that, unlike the case of static users \cite{banerjee2020fundamental}, there could be more than one successful transmissions within a super-interval by users other than the \textsf{Max}-user.   
We now claim that the \textsf{Max}-user corresponding to the $i$\textsuperscript{th} super-interval must have a successful transmission by the beginning of the last $N-1$ super-intervals. If not, by the pigeonhole principle, some other user $j\neq M_i$ must be the \textsf{Max}-user at least twice in the previous $N$ super-intervals. However, this cannot be true as the user $j$ would have less age than $M_i$ when the user $j$ became the \textsf{Max}-user for the second time in the previous $N$ super-intervals.     

	
	Hence, at the $k$\textsuperscript{th} slot of the $i$\textsuperscript{th} super-interval, the age of the \textsf{Max}-user $M_i$ is upper bounded by $k+ \sum_{j=1}^{N-1} \Delta_{i-j},$ where for notational consistency, we have defined $T_j \equiv 0, \textrm{and } \Delta_j \equiv 0, \forall j\leq 0.$
 Thus, the cost $C_i^{\textsf{CMA}}$ incurred by the \textsf{CMA} policy during the $i$\textsuperscript{th} interval may be upper-bounded as: 
 \begin{eqnarray} \label{CMA}
	C_i^{\textsf{CMA}} &\leq& \sum_{k=1}^{\Delta_i} \bigg(k+ \sum_{j=1}^{N-1} \Delta_{i-j}\bigg)
	= \frac{1}{2}\big(\Delta_i^2+\Delta_i) + \sum_{j=1}^{N-1}\Delta_i \Delta_{i-j} \nonumber\\
	&\leq& \frac{1}{2}\big(\Delta_i^2+\Delta_i) + \frac{1}{2}\sum_{j=1}^{N-1}\big(\Delta_i^2 + \Delta_{i-j}^2\big) \label{am_gm}\\
	&=& \frac{N}{2}\Delta_i^2 + \frac{1}{2}\Delta_i + \frac{1}{2}\sum_{j=1}^{N-1} \Delta^2_{i-j}. \nonumber
 \end{eqnarray}
	where in Eqn.\ \eqref{am_gm}, we have used the AM-GM inequality to conclude $ \Delta_i \Delta_{i-j} \leq \frac{1}{2}\big(\Delta_i^2 + \Delta_{i-j}^2\big), 1\leq j \leq N-1.$
	Hence, assuming that there are a total of $K$ super-intervals in the time-horizon $T$, the total cost incurred  by the \textsf{CMA} policy over the entire time horizon is upper bounded as: 
	\begin{eqnarray*}
	\textrm{AoI}^{\textsf{CMA}}(T)&=& \sum_{i=1}^{K}C_i^{\textsf{CMA}} 
	\leq  \frac{1}{2}\sum_{i=1}^{K} \bigg(2N \Delta_i^2 + \Delta_i\bigg).
	\end{eqnarray*}
	
On the other hand, the cost (\emph{i.e.,} the maximum age among all users) incurred by \textsf{OPT} during the $i$\textsuperscript{th} super-interval is trivially lower bounded by the age of the user $M_i$, which was consistently experiencing \textsf{Bad} channels throughout the $i$\textsuperscript{th} super-interval, \emph{i.e.,}
	\begin{eqnarray} \label{COPT}
	C_i^{\textsf{OPT}}\geq  \sum_{k=1}^{\Delta_i} (1+k)
	=  \frac{1}{2} \Delta_i^2 + \frac{3}{2}\Delta_i, 
	\end{eqnarray}

Finally, the cost of the entire horizon may be obtained by summing up the cost incurred in the constituent intervals. 
	Hence, noting that $\Delta_0=0$, from Eqns.\ \eqref{CMA} and \eqref{COPT}, the competitive ratio $\eta^{\textsf{MA}}$ of the \textsf{CMA} policy may be upper bounded as follows:
	\begin{eqnarray*}
	\eta^{\textsf{CMA}} = \frac{\sum_{i=1}^K C_i^{\textsf{CMA}}}{\sum_{i=1}^K C_i^{\textsf{OPT}}}	
	\stackrel{(a)}{\leq} \frac{\frac{1}{2}\sum_{i=1}^{K} \bigg(2N \Delta_i^2 + \Delta_i\bigg)}{\sum_{i=1}^K \big(\frac{1}{2}\Delta_i^2 + \frac{3}{2}\Delta_i\big)} 
	\leq 2N.~~ \blacksquare
	\end{eqnarray*}

\section{Converse} \label{lower_bound_section}
\begin{theorem} \label{lower_bound}
For any online policy $\mathcal{A}$, $\eta^{\mathcal{A}}\geq \Omega(\frac{N}{\ln N}).$	
\end{theorem}
 
\textsc{Proof:}
We establish a slightly stronger result by proving the lower bound for the particular case when all $N$ users remain stationary at a single cell throughout the entire time interval. 
Using Yao's minimax principle, a lower bound to the competitive ratios of \emph{all} deterministic online algorithms under \emph{any} input channel state distribution $\bm p$ yields a lower bound to the competitive ratio, \emph{i.e.,} 
\begin{eqnarray}\label{Yao_lb}
\eta \geq \frac{\mathbb{E}_{\bm{\sigma} \sim \bm{p}}(\textrm{Cost of the Best Deterministic Online Policy})}{\mathbb{E}_{\bm \sigma \sim \bm p}\textrm{(Cost of OPT)}}.	
\end{eqnarray}
To apply Yao's principle in our setting, we construct the following symmetric channel state distribution $\bm{p}$:  at every slot $t$, a user is chosen independently and uniformly at random and assigned a \textsf{Good} channel. The rest of the $N-1$ users are assigned \textsf{Bad} channels. 
 Hence, at any slot $t$:
$\mathbb{P}(\textrm{user}_i\textrm{'s channel is \textsf{Good}})=\frac{1}{N},$ and is \textsf{Bad} otherwise. 
The rationale behind the above choice of the channel state distributions will become apparent when we compute \textsf{OPT}'s expected cost below. In general, the cost of the optimal offline policy is obtained by solving a Dynamic Program, which is challenging to analyze. However, with our chosen channel distribution $\bm{p}$, we see that only one user's channel is in \textsf{Good} state at any slot. This greatly simplifies the computation of \textsf{OPT}'s expected cost.  
We lower bound the competitive ratio using Eqn.\ \eqref{Yao_lb} by lower bounding the numerator and upper bounding the denominator for the symmetric channel state distribution described above. 

\textbf{An Upper bound to \textsf{OPT}'s cost:} The \textsf{OPT} policy, with a priori channel state information, schedules the only user having a \textsf{Good} channel at any slot. Hence, the limiting distribution of the age of any user is Geometric ($\frac{1}{N}$), \emph{i.e.,}
\begin{eqnarray*}
\lim_{t \to \infty} \mathbb{P}(h_i(t)=k) = \frac{1}{N}\big(1-\frac{1}{N}\big)^{k-1}, ~~ k \geq 1.	
\end{eqnarray*}
Hence, for upper bounding the time-averaged cost incurred by the \textsf{OPT} policy, using Cesaro's summation formula, it is enough to  upper bound the expected value of maximum of $N$ \emph{dependent} but identically Geometrically distributed random variables.
The MGF of the Geometric distribution $G$ is: 
\begin{eqnarray*}
\mathbb{E}(\exp(\lambda G)) =\begin{cases}
 	\frac{e^\lambda/N}{1-e^\lambda(1-1/N)},~~\textrm{ if } \lambda < -\log(1-1/N)\\
 	\infty ~~ \textrm{o.w.}
 \end{cases}
\end{eqnarray*}
Let the r.v.\ $H_{\max}$ denote limiting maximum age of the users. 
We proceed similarly to the proof of Massart's lemma for upper bounding $\mathbb{E}(H_{\max}).$ For any $-\log(1-1/N)>\lambda >0,$ we have
\begin{eqnarray*}
	&&\exp\big(\lambda \mathbb{E}(H_{\max})\big)\\
	&\stackrel{(a)}{\leq}&  \mathbb{E}(\exp(\lambda H_{\max})) 
	\leq \sum_{i=1}^N \mathbb{E}(\exp(\lambda G_i)) 
	\leq \frac{e^\lambda}{1- e^\lambda (1-\frac{1}{N})},
\end{eqnarray*}
where the inequality (a) follows from Jensen's inequality.
Taking natural logarithm of both sides, we get
\begin{eqnarray}\label{H_max_bd}
 \mathbb{E}(H_{\max}) \leq 1 - \frac{1}{\lambda}\log\big(1-e^\lambda (1-1/N)\big).	
\end{eqnarray}
Now, let us choose $\lambda= \frac{\alpha}{N},$ for some fixed $0< \alpha <1$ to be determined later. First, we verify that, with this choice for $\lambda$, we always have $\lambda < -\log(1-\frac{1}{N})$. Using the convexity of the function $e^x,$ we can write
\begin{eqnarray}\label{cvx_bd}
1=e^{0} \geq e^x + (0-x) e^x = (1-x)e^x \implies  e^x \leq \frac{1}{1-x}, x<1.
\end{eqnarray}
As a result, we have 
\begin{eqnarray*}
e^\lambda\equiv e^{\frac{\alpha}{N}} \leq \frac{1}{1-\frac{\alpha}{N}} < \frac{1}{1-\frac{1}{N}}; \textrm{ i.e., } \lambda < -\log(1-\frac{1}{N}). 	
\end{eqnarray*}
Next, for upper bounding the RHS of Eqn.\ \eqref{H_max_bd}, we start with the simple analytical fact that for $0<\alpha <1,$
\begin{eqnarray}\label{analysis1}
	\inf_{0<x<1} \frac{1-(1-x)e^{\alpha x}}{x} = 1-\alpha.
\end{eqnarray}
This result can be verified by using Eqn.\ \eqref{cvx_bd} to conclude that for $0<x<1,$ we have
\begin{eqnarray*}
	\frac{1-(1-x)e^{\alpha x}}{x} \geq \frac{1}{x}\big(1- \frac{1-x}{1-\alpha x}\big)= \frac{1-\alpha}{1-\alpha x} \geq 1-\alpha,
\end{eqnarray*}
where the infimum is achieved when $x\to 0^+.$ Substituting $x =\frac{1}{N}$ in the inequality \eqref{analysis1}, we have the following bound 
\begin{eqnarray*}
1-e^{\alpha/N}(1-1/N)\geq \frac{1-\alpha}{N}.	
\end{eqnarray*}
Hence, using Eqn.\ \eqref{H_max_bd}, we have the following upper bound to the expected Max-age under \textsf{OPT}:
\begin{eqnarray*}
\mathbb{E}(H_{\max})\leq 1+ \frac{N}{\alpha}\ln \frac{N}{1-\alpha},	
\end{eqnarray*}
for some $0< \alpha <1.$ 
Setting $\alpha = 1-\frac{1}{\ln N}$ yields the following asymptotic bound: 
\begin{eqnarray*}
	\mathbb{E}(H_{\max}) \leq N\ln N + o(N\ln N). 
\end{eqnarray*}
\textbf{Lower Bound to the cost of any online policy:}
To lower bound the cost of any online policy $\mathcal{A}$, we use Theorem 1 of \cite{srivastava2019minimizing} with the success probability $p_i=\frac{1}{N}, \forall i,$ yielding:
\begin{eqnarray*}
	\liminf_{T \to \infty} \frac{1}{T}\sum_{t=1}^T \mathbb{E}(\max_i h^{\mathcal{A}}_i(t)) \geq N^2.
\end{eqnarray*}
Combining the above results and using Eqn.\ \eqref{Yao_lb}, the competitive ratio of any online algorithm is lower bounded as: 
\begin{eqnarray*}
\eta^{\mathcal{A}}  \geq \sup_{T >0} \frac{C^{\pi}(T)}{C^{\textsf{OPT}}(T)} \geq \Omega(\frac{N}{\ln N}).~~\blacksquare
\end{eqnarray*}

\bibliographystyle{unsrt}
\bibliography{ref-MAMA}

\end{document}